\documentclass[sigconf,authorversion,nonacm]{acmart}

\setcopyright{none}

\usepackage{wrapfig}
\usepackage{amsmath}
\usepackage{mathtools}
\usepackage{algorithm}
\usepackage{algpseudocode}
\usepackage{listings}
\usepackage{epsfig}
\usepackage{graphicx}
\usepackage{subcaption}
\usepackage{hyperref}
\usepackage{breakurl} 
\usepackage{mathpartir}
\usepackage{mathtools}
\usepackage{wrapfig}
\usepackage{enumitem}
\usepackage{verbatimbox}
\usepackage{multirow}
\usepackage{lipsum}
\usepackage{caption}
\usepackage{todonotes}
\usepackage{adjustbox}
\usepackage{listings, xcolor}
\newenvironment{nop}{}{}

\newcommand{\tool}{{\sc ClassFinder}}
\newcommand{\Q}{{\tt Q}}
\newcommand{\R}{{\tt R}}

\newcommand{\rulesep}{\unskip\ \vrule\ }
\begin{document}
\title{Searching for Replacement Classes}

\author{Malavika Samak}
\email{malavika@csail.mit.edu}
  \affiliation{%
    \institution{CSAIL, MIT}
    \country{USA}
   }
    
\author{Jose Pablo Cambronero}
 \email{jcasman@csail.mit.edu}
  \affiliation{%
    \institution{CSAIL, MIT}
    \country{USA}
  }
 
 \author{Martin C. Rinard}
\email{rinard@csail.mit.edu}
  \affiliation{%
    \institution{CSAIL, MIT}
    \country{USA}
}


%
%


\sloppy

\begin{abstract}
  Software developers must often replace existing components in their systems to
  adapt to evolving environments or tooling. While traditional code search
  systems are effective at retrieving components with \emph{related}
  functionality, it is much more challenging to retrieve components that can be
  used to directly replace existing functionality, as replacements must account
  for more fundamental program properties such as type compatibility. To address
  this problem, we introduce \tool{}, a system which given a query class Q, and
  a search corpus S, returns a ranked subset of classes that can replace Q and
  its functionality. \tool{} produces a field and method mapping between the
  classes that can provide useful hints to a developer and can be used to
  effectively refine the ranking of candidate replacement classes. Our technique
  leverages the complementary strengths of a distributed embeddings-based search
  and type-based analysis, using the former to prune down candidates for an
  optimization-based approach based on the latter. \tool{} retrieves replacement
  classes, along with a type-aware field/method mapping between classes. We
  evaluate \tool{} on a search space of $\approx 600$ thousand open source Java
  classes. Querying \tool{} with 24 known Java classes provided meaningful
  replacement classes and mappings, in many cases producing complete mappings
  with functionally-identical replacement classes.

\end{abstract}

\maketitle

\section{Introduction}
Developers maintaining software systems often need to replace one of the classes in the source code base with a functionally equivalent class. Potential motivations include deprecation of deployed classes~\cite{deprecatedjdk,classdeprecation,deprecated-react}, the need to change vendors to satisfy organizational needs or intellectual property constraints~\cite{oraclevsgoogle,guavamotivation}, improved performance and/or memory usage~\cite{guavaapache,energydiff}, or the desire to identify and use better versions that may contain fewer defects~\cite{evosuite,jdkbugfixes,java-security}. While large software repositories often contain suitable class replacements, finding such replacements can be a challenging task given the enormous size of extant code bases (billions of lines of code)~\cite{github100million,codegrowth}.

We present a new technique and implemented system, ClassFinder, for automatically finding replacement classes. Given a query class, ClassFinder automatically searches large code bases to identify and rank potential replacement classes. ClassFinder combines two complementary techniques: embedding-based class ranking and method compatability matching. Treating the method names of each class as a bag of words, embedding-based class ranking maps the names to a high-dimensional vector, with the cosine similarity metric over the resulting vector space measuring the distance between classes. We find that, in practice, this metric provides a reasonably accurate proxy for the intended semantic similarity between the query class and candidate replacement classes. However, because it ignores method type and implementation information, it fails to adequately capture aspects relevant to class replacement\cite{mask}.

ClassFinder therefore next deploys a compatability matching algorithm that operates at the level of individual methods. Working with the tokens in each method as a bag of words, the matching computes an embedding for each method, with the cosine metric used as a proxy for semantic method similarity. Matching methods in the original and candidate replacement classes according to this metric, ClassFinder computes a type-similarity metric between matched methods (Section ~\ref{sec:method-mapper}). ClassFinder also matches fields in the original and candidate classes to compute a field mapping metric (Section ~\ref{sec:field-mapper}). Aggregating these metrics enables ClassFinder to effectively identify and rank candidate replacement classes. In particular, we find that compatability matching is effective at deranking candidate replacement classes that 1) operate with incompatible or overly specific types, 2) contain empty or placeholder methods, or 3) implement a subset of the required methods.  

We evaluate \tool{} on {\tt 24} open source
{\tt Java} query classes. These classes contained deprecated classes, classes used by Samak et al\cite{mask} and classes from popular packages\cite{java-popular}. Working with a corpus of $\approx600$ thousand open-source candidate replacement classes, our results indicate that ClassFinder can effectively find suitable replacement classes for our set of query classes within this large candidate replacement class search space. Specifically, ClassFinder found replacement classes that
provide {\em complete} replacement to the query class as the top candidate for 12 of the query classes and partial replacements ranging between 30\% to 95\% replacement for the rest (Section~\ref{sec:experiments}). The results also indicate that the embedding-based
search used in the first phase of our approach identifies classes that
were good replacements as the top candidate in some cases, but identifies type incompatible or otherwise unsuitable replacements in others, with the combination of embedding-based search and compatability matching significantly outperforms embedding-based search alone. 

\begin{figure*}
  \begin{subfigure}[t]{.24\linewidth}
    \begin{lstlisting}[language = Java , firstnumber = last , escapeinside={(*@}{@*)}]
public class ArrayList<E> {
  transient Object[] elementdata; 
  int size;
  ...
  boolean contains(Object o) {
    return indexOf(o) >= 0;
  }
      
  E set(int i, E element) {
    Objects.checkIndex(i, size);
    E old = elementData(i);
    elementData[i] = element;
    return old;
  }
} 
\end{lstlisting}\subcaption{}
\end{subfigure}
\rulesep
\begin{subfigure}[t]{.23\linewidth}
\begin{lstlisting}[language = Java , firstnumber = last , escapeinside={(*@}{@*)}]
public class ImmutableList<E> {
  final E[] arr;
  
  boolean contains(Object o) {
    return indexOf(o) != -1;
  }
  ...
  E set(int i, E e) { 
    throw new UOE();
  }
} 
\end{lstlisting}\vspace{0.45in}\subcaption{}
\end{subfigure}
\rulesep
\begin{subfigure}[t]{.24\linewidth}
\begin{lstlisting}[language = Java , firstnumber = last , escapeinside={(*@}{@*)}]
public class BooleanList {
  boolean[] elementData;
  int size = 0;
  
  boolean contains(boolean e) {
    return indexOf(e) >= 0;
  }
  ...
  boolean set(int i, boolean e) {
    rangeCheck(i);
    boolean old = elementData[i];
    elementData[i] = e;
    return old;
  }
} 
\end{lstlisting}\subcaption{}
\end{subfigure}
\rulesep
\begin{subfigure}[t]{.24\linewidth}
\begin{lstlisting}[language = Java , firstnumber = last , escapeinside={(*@}{@*)}]
public class Vector<E> {
  transient Object[] elementdata;
  int elementCount;
  
  boolean contains(Object o) {
    return indexOf(o) >= 0;
  }
  ...
  E set(int i, E element) {
    if (i >= elementCount) 
      throw new OOBE(i);
    E old = elementData(i);
    elementData[i] = element;
    return old;
  }
} 
\end{lstlisting}\vspace{-0.1in}\subcaption{}
 \end{subfigure}
  \caption{The {\tt ArrayList} implementation from {\tt java.util} package and its potential replacement candidates: {\tt ImmutableList} from  {\tt jadax.core}, {\tt BooleanList} from {\tt abacus.util} and {\tt Vector} from {\tt java.util}.} \label{fig:arraylist-repl}
\end{figure*}


This paper makes the following contributions:
\begin{itemize}[leftmargin=*]
    \item {\bf Problem and Technique:} It identifies the replacement class search problem and presents a technique and system for solving this problem. The technique uses {\em embedding-based class ranking} to efficiently identify candidate replacement classes with good semantic similarity to the query class. It then uses method-based compatibility matching to identify candidate replacement classes whose methods exhibit good type and semantic compatibility with the query class.
    \item {\bf System:} It presents the design and implementation of ClassFinder, a system that implements our class search technique. 
    \item {\bf Results:} It presents results that characterize the effectiveness of ClassFinder in finding replacement classes for 24 query classes within a corpus of $\approx600$ thousand candidate open-source replacement classes. These results characterize the synergistic combination of embedding-based class ranking and method-based compatability matching, both of which are required for ClassFinder to effectively identify appropriate replacement classes for our benchmark query classes. 
\end{itemize}

\section{Class Replacement Example}\label{sec:example}
In this section, we explain the working of \tool{} with an example.
Consider the {\tt ArrayList} class from 
{\tt java.util} package shown in Figure~\ref{fig:arraylist-repl}.
The class provides a resizable implementation of a list and it  can be parameterized to create and operate on a list of any reference type {\tt E}.  We search for replacements to this class within a search space of over 600 thousand classes. Among many other classes, this search space contains 
{\tt ImmutableList} from {\tt jadax.core}, {\tt BooleanList} from {\tt abacus.util} and {\tt Vector} from {\tt java.util} package shown in Figure~\ref{fig:arraylist-repl}.

The goal of the search is to deliver a ranked list of classes that offer a complete or partial replacement to {\tt ArrayList}, where the rank is determined based on the number of  
drop-in replacements to {\tt ArrayList} methods. A class that offers partial replacement may still be useful if the developer is either willing to  write adapter methods
for the unmapped methods or plans to use a subset of the methods offered by the query class. 
All three classes from Figure~\ref{fig:arraylist-repl} can provide {\em at-least} a partial replacement to {\tt ArrayList}, therefore are relevant to the search. 

To identify these classes from our corpus of classes, 
\tool{} initiates a search for classes similar to {\tt ArrayList}. \tool{} bases this search on the assumption: {\em method 
names are good descriptor of the functionality}. 
It extracts the
method names defined by the class, creates a bag of words containing the method names, then maps the bag of words into a high dimensional vector space~\cite{bojanowski2016enriching}. It then queries the set of corresponding vectors from all of the classes in the search space to find the closest classes to the query class as measured by the cosine distance between vectors~\cite{cbow}. Reflecting the fact that they implement similar functionality as {\tt ArrayList}, in our example the vector embeddings of {\tt ImmutableList}, {\tt BooleanList}, and {\tt Vector} all exhibit small cosine distances from the vector embedding of {\tt ArrayList} and these three classes appear at the top of the embedding ranking. 

If the embedding based rank was used as the final result, all three classes would be identified as equally good replacements to {\tt ArrayList}. But a closer examination of the classes reveals important differences. 
The {\tt Vector} class offers
a complete replacement to {\tt ArrayList}, whereas {\tt ImmutableList} offers a partial
replacement because it can not replace {\tt set} and other methods that modify the list. Similarly {\tt BooleanList} can not offer complete replacement either, as it narrows the
underlying type of the list to {\tt boolean} and therefore supports only {\tt boolean} lists.

\tool{} next matches methods defined by {\tt ArrayList} to those defined by each of the candidate classes. The match is based on the type signature and the implementation of each method. 
The result of these method matches determines the rank of each candidate class. The functional equivalence of two methods is measured by creating an embedding based on their implementation and computing the cosine similarity between them. \tool{} 
computes the similarity between all pairs of methods from {\tt ArrayList} and candidate class. It also evaluates
the type compatibility between
all pairs of methods.

Given a method from {\tt ArrayList}, \tool{} matches it to the method in each candidate 
class that is type compatible and is most similar to the {\tt ArrayList} method.
If no such candidate can be found, the method will be matched to $\bot$. \tool{} also ensures the similarity score 
for a pair of matched methods is above a threshold set by the user. Otherwise the query class method will be matched to $\bot$. 
In our example, the {\tt set} method in {\tt ArrayList} will be matched to the {\tt set} method in {\tt Vector}, but will be matched to $\bot$ for other two classes --- there are no type compatible methods in {\tt BooleanList} and no high similarity methods in {\tt ImmutableList}. 
The {\tt contains} method in the {\tt ArrayList} is similarly matched. Finally, we see that the method matches for {\tt Vector} contain drop-in replacements for both methods, whereas {\tt ImmutableList} contains a match
for the {\tt contains} method only and {\tt BooleanList} does not contain any matches because of type incompatibility.  The resulting ranking correctly ranks {\tt Vector}
as the top candidate, followed by {\tt ImmutableList} and
finally {\tt BooleanList}.

All through this section we have assumed that, the type compatibility and functional
equivalence are equally important to the user. We recognize this need not always be the case. 
\tool{} allows the user to set a higher weight to type compatibility or
functional similarity with a configuration variable. Setting a higher weight for one of them can change the final rank. For example, if
zero weight was set for type compatibility, {\tt BooleanList} would rank above {\tt ImmutableList}. 

\section{Design}\label{sec:design}
\begin{figure}[h]
    \begin{center}
        \includegraphics[width=0.5\textwidth]{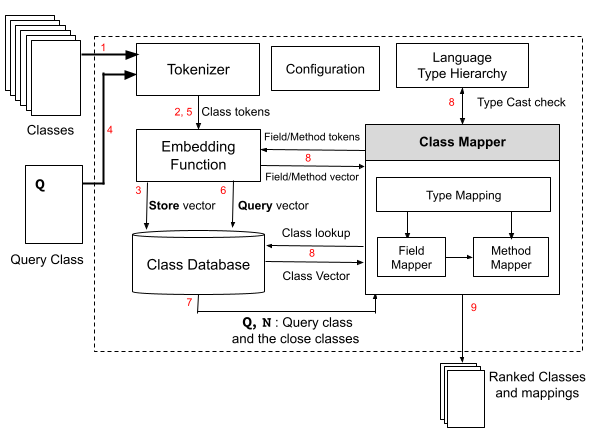}
        \caption{\vspace{-0.1in}The \tool{} architecture}
        \label{fig:architecture}
    \end{center}
\end{figure}

Figure~\ref{fig:architecture} presents the overall design of \tool{}. \tool{}
pre-processes every class in its search space and creates a class search database.
Each class available to the tool is first tokenized by
extracting the field and method names in its implementation.
The tokens extracted from the class are input to an embedding function, which
computes a corresponding vector. This vector is then stored in our database.
When queried with a class {\tt Q}, \tool{} similarly computes a vector
for the query class and retrieves a list of
the most similar candidate classes {\tt N}, based on the cosine similarity of their vectors.
This step is designed to identify a smaller set of candidate classes before moving
on to more expensive analyses.

Every class {\tt R} in the set {\tt N} is analyzed by the {\tt Class Mapper}, which constructs a field map $\sigma$ and a method map $\alpha$
between {\tt Q}
and {\tt R}. All classes in {\tt N} are re-ranked based on these newly constructed maps.
The field map is constructed by assigning a score to every pair of fields in \Q{} and \R{}. The score is
based on the type compatibility between fields and the vector distance between their embeddings. Depending on the data types, the compatibility is inferred either by using the language's {\em Type Hierarchy} (which handles primitive data types, classes offered by the language, etc) or based on their
type {\em similarity} computed based on the cosine similarity between the corresponding vectors stored in the database. 
Similarly, the method map $\alpha$ is constructed
based on the type signatures of the methods and the vector embedding of the method body. The list of
re-ranked classes and their method maps is output to the user. 
\subsection{\tool{}: Top-level algorithm}
\begin{algorithm}[h]
	\footnotesize
	\caption{\label{alg:overall} \tool{} main algorithm}
	\begin{algorithmic}[1]
		\Require{\Statex \Q{} $\gets$ Query class}
        \State Initialize $\mathtt{DB}$ with embeddings of all available classes
		\State $\mathtt{tokens \gets ClassTokenizer(\Q{})}$;
		$\mathtt{vec \gets embedding(queryTokens)}$
		\State $\mathtt{N \gets lookup(DB, vec)}$
		\For{every $\mathtt{R \in N}$} \Comment{Class mapping}
		\State ${TS} \gets \mathtt{TypeSimilarity(\texttt{Q, R})}$
		\State $\sigma \gets \mathtt{FieldMapper}(\texttt{Q, R, } TS)$;
 		\State $\alpha \gets \mathtt{MethodMapper}(\texttt{Q, R, }\sigma,~{TS})$
		\State $\mathtt{result[R]} \gets \langle \sigma, \alpha \rangle$
		\EndFor
		\State  \textbf{return} $\mathtt{rank(result)}$
	\end{algorithmic}
\end{algorithm}

Algorithm~\ref{alg:overall} presents the top-level algorithm of \tool{}.
Initially, every class in the search corpus is tokenized, the extracted tokens are  embedded into vectors, and which are then stored in the database.
After initializing the database, the \tool{} can be queried with a
class {\tt Q}. The input query class \Q{} is
tokenized and embedded into a vector at line {\tt 2}.
The resulting vector is used to search the class database {\tt DB} for {\tt N}, a list of
classes in the database close to the query class, sorted based on
their vector similarity.

The algorithm iterates over every class {\tt R} contained in {\tt N} ({\tt line 4-9}). For a given (\Q{}, {\tt R}) pair, the algorithm creates a type similarity matrix ${TS}$, with the help of
	{\tt TypeSimilarity} procedure. The matrix captures the similarity between a pair of data
types ($\mathtt{T_i, T_j}$), where type $\mathtt{T_i}$ is used by class {\tt Q}  and
type $\mathtt{T_j}$ is used by class {\tt R}.
The ${TS}$ matrix assigns a score to each pair ($\mathtt{T_i, T_j}$), which indicates if
type $\mathtt{T_i}$ can be mapped to type $\mathtt{T_j}$. The {\tt FieldMapper} creates embeddings for all fields in class \Q{} and {\tt R}. The embedding based similarity between
a pair of fields and the type similarity score in $TS$ guide the creation of field map $\sigma$ which is used by the {\tt MethodMapper} to rewrite the class {\tt R}.
In particular, fields are renamed based on $\sigma$ to account for possible naming
differences between methods. The {\tt MethodMapper} then creates embedding
for all methods defined by class \Q{} and {\tt R} and computes the vector similarity for
all pairs of methods. This vector similarity and the type similarity from $TS$
are used to create a method map $\alpha$. The constructed $\sigma,\alpha$ pair is added
to the {\tt result} map. Finally, the entries in {\tt result} are ranked and output to user.

\subsection{Continuous Representations of Code}\label{sec:embedder}
The $\mathtt{ClassTokenizer}$ receives a class $\mathtt{C}$ as input and returns a bag 
of words containing the name of the class, names of all the fields defined by the class, 
the name of every method defined by the class and their formal parameter names. The names 
of methods invoked by the class are also added to this list. Camel cases and underscores are processed to split combined words. The resulting sequence is returned to the caller. For the {\tt ArrayList} code shown in Figure~\ref{fig:arraylist-repl} the $\mathtt{ClassTokenizer}$ returns the following: {\tt array}, {\tt list}, {\tt element}, {\tt data}, {\tt size}, {\tt contains}, {\tt o}, {\tt index}, {\tt of}, {\tt set}, {\tt i} and {\tt check}.

Then, to identify candidate replacement classes, \tool{} makes use of an \emph{embedding} function that
can compute a vector representation for a given class. This embedding
function, $F: T \rightarrow \mathbb{R}^d$, takes a sequence of tokens ($T$)
extracted from a code snippet and computes a corresponding real-valued vector
of dimension $d$. The goal of this embedding is to capture semantic properties
of the code fragment. Given two sequence of tokens, \tool{} applies $F$ to each sequence, and then computes a similarity
score between the two resulting vectors. \tool{}, like related code search
systems~\cite{sachdev2018retrieval, gu2018deep, cambronero2019deep}, uses cosine
similarity (which ranges between -1 and 1) as the similarity function over two
vectors, such that vectors that are similar have a score closer to 1.

In our implementation, we instantiate $F$ by training a continuous-bag-of-words
(CBOW)~\cite{cbow} model using FastText~\cite{bojanowski2016enriching}. In a
CBOW model, we take a sequence of tokens $t_1, ..., t_n$ and learn
representations such that we can predict the presence of a token $t_i$ given a
representation for the context (i.e. the remaining tokens in the sequence). In
practice, CBOW models take the context to be the tokens that occur within a
(potentially randomly sampled-sized) window of the target token $t_i$.  In
FastText, a vector representation for the context is computed by first
extracting n-gram subwords (subsequences at the character level) from the
corresponding sequence of tokens, mapping each such subword to its own
vector, and then averaging the subword vectors (i.e. mean-pooling). Critically,
the use of subwords allows this CBOW model to produce vectors even for tokens
that are otherwise out-of-vocabulary with respect to the set of tokens observed
during training. When training, standard backpropagation is used to update the
subword embeddings used in the model. We train the CBOW model, with embeddings
of dimension 150, on a corpus of ~600K classes. For more details on CBOW models, we refer the reader to
Mikolov et al~\cite{cbow}.

\subsection{Type Similarity Matrix} 
Determining whether type $\mathtt{T_i}$ in class \Q{} can be mapped to a value of
type $\mathtt{T_j}$ in class {\tt R} is crucial for the overall success of \tool{}. 
This knowledge guides the creation of field and method mappings, which are 
used to assess if class {\tt R} is a good replacement for \Q{}.
The language's type hierarchy provides a definite yes/no answer to this
question for primitive types and
classes that inherit from one another.
However, if the types compared are unrelated,
\tool{} checks if $\mathtt{T_i}$ is similar to $\mathtt{T_j}$ based on
their embedding similarity. 
If they have a high similarity, \tool{} determines that the types can be mapped. 
 
\tool{} uses the $\mathtt{TypeSimilarity}$ procedure presented in Algorithm~\ref{alg:typescore} to assign
a similarity score for all type pairs and stores the scores in matrix ${TS}$.
\begin{algorithm}[h]
	\footnotesize
	\caption{\label{alg:typescore} $\mathtt{TypeSimilarity}$ algorithm}
	\begin{algorithmic}[1]
		\Require{\Statex {\tt Q} $\gets$ Query class; {\tt R} $\gets$ Replacement class;} 
		\Statex $\mathtt{tt} \gets$ Type similarity threshold\Comment{Configuration variable}
		\State ${TS} \gets \mathtt{INIT(Q, R)}$;
		 $\mathtt{TS[\texttt{Q,R}]} \gets \mathtt{1}$ 
		\For {every $\mathtt{(T_i, T_j)} \in \mathtt{Types(Q)} \times \mathtt{Types(R)}$}
		\If{$\mathtt{typeCastCheck(T_i, T_j)}$}
		$TS\mathtt{[T_i,T_j] \gets  \mathtt{1}}$
		\ElsIf{$\mathtt{T_i, T_j}$ are classes}\Comment{Embedding based type similarity}
		\State $\mathtt{vec}_{1} \gets \mathtt{lookup(DB, T_i)}$;
		~$\mathtt{vec}_{2} \gets \mathtt{lookup(DB, T_j)}$
		\State $\mathtt{score} \gets \mathtt{similarity(vec_1, vec_2)}$
		\If{$\mathtt{score} > \mathtt{tt}$}
		$TS\mathtt{[T_i,T_j] \gets
				score}$
		\EndIf
 		\EndIf
		\EndFor
		\State \textbf{return} ${TS}$
	\end{algorithmic}
\end{algorithm}
The algorithm takes a query class {\tt Q} and a replacement
class {\tt R} as inputs and assigns a score
to every type pair ($\mathtt{T_i,T_j}$) used by \Q{} and
	{\tt R}. The scores range between {\tt -1 to +1}.
{\tt +1} indicates that $\mathtt{T_i}$ can be mapped to type $\mathtt{T_j}$ and {\tt -1} indicates mapping
is not permitted. The algorithm initializes all entries to {\tt -1} with the {\tt INIT} function.
\tool{} sets {\tt +1} as the similarity score for the ({\tt Q,R}) pair. This allows
mapping instances of type \Q{} to instances of type {\tt R} as a starting point.
For every other pair of types $\mathtt{(T_i, T_j)}$, the
type score is calculated (lines {\tt 2-10}). If type
$\mathtt{T_i}$ can be cast to $\mathtt{T_j}$, a maximum
replacement score {\tt 1} is set for $\mathtt{(T_i, T_j)}$. Otherwise, if both are non-primitive types, the embedding based similarity score is computed for
the pair (lines {\tt 4-9}). 
The class vectors for $\mathtt{T_i,T_j}$ is and their cosine similarity score is computed by the {\tt similarity}. 
If the score is above a user defined threshold $\mathtt{tt}$, the score is added to the matrix (line 7). 
Otherwise, the original {\tt -1} score is maintained.
\subsection{Field Mapper}\label{sec:field-mapper}
{\tt FieldMapper} attempts to map every field {\tt f} in class \Q{}, to a field {\tt g} in class {\tt R}, by
considering type similarity and the field uses within \Q{} and {\tt R}. It identifies the set of
methods in the class that read or write a given field.
The fields that are read and written by a similar set of methods are given a higher similarity score. 
This score along with the type similarity
scores is used to create the field mapping, $\sigma$.
Algorithm~\ref{alg:fieldScore} presents the overall working of {\tt FieldMapper}.
\begin{algorithm}[h]
	\footnotesize
	\caption{\label{alg:fieldScore} $\mathtt{Field-Mapper}$ algorithm}
	\begin{algorithmic}[1]
		\Require{\Statex {\tt Q} $\gets$ Query class; {\tt R} $\gets$ Replacement class; $TS$ $\gets$ Type similarity matrix
		\Statex $\mathtt{ft} \gets$ Field similarity threshold \Comment{Configuration variables}
		\Statex$\mathtt{fw} \gets$ Field embedding weight
		}
		\State ${FS} \gets \mathtt{INIT(Q,R)}$
		\For {every $\mathtt{(f, g)} \in \mathtt{Fields(\texttt{Q})} \times \mathtt{Fields(\texttt{R})}$}
		\If{{\tt modifier-compatibility(f,g) = false}} \textbf{continue};
		\EndIf
		\State $\langle \mathtt{rToken_f, wToken_f} \rangle\gets \mathtt{fieldReadWrites(f, Q)}$; \Comment{Read, write tokens}
		\State $\langle \mathtt{rToken_g, rToken_g} \rangle\gets \mathtt{fieldReadWrites(g, R)}$; 
        \State $\mathtt{rScore} \gets \mathtt{similarity(rToken_f,rToken_g)}$\Comment{Read, write similarity score}
        \State $\mathtt{wScore} \gets \mathtt{similarity(wToken_f,wToken_g)}$
		\State $\mathtt{escore} \gets \mathtt{(rScore+wScore)}/2$
		\State ${FS[{\tt f, g}]} \gets$
		$\mathtt{fw} * \mathtt{escore} + (1 -\mathtt{fw})  * {TS[{\tt f, g}]}$ \Comment{Similarity score for {\tt f,g}}
		\EndFor
    	\State $\mathtt{field\text{-}map} \gets \mathtt{OPTIMIZE}(FS)$
    	\State $\sigma \gets \mathtt{filter}(\mathtt{field\text{-}map}, \mathtt{ft})$
		\State \textbf{return} $\sigma$
	\end{algorithmic}
\end{algorithm}

Initially, ${FS}$ is initialized by assigning a {\tt -1} score for all field pairs. The algorithm iterates over every combination of fields $\mathtt{(f, g)}$
 defined by \Q{} and {\tt R}. The modifiers attached
 to the fields {\tt f,g} are first checked for compatibility with {\tt modifier-compatibility} check. If the field modifiers are compatible, then the check returns {\tt true}. If the check returns {\tt false}, 
 the pair is discarded and the next pair is processed (line {\tt 3}). This check helps to
 eliminate erroneous maps, where a non-static field is mapped to
 a static field or a mutable field is mapped to an immutable field, etc.  
 
 For every compatible field pair ({\tt f,g}), the similarity
 between their read and write uses is computed. To compute the read/write similarity, the set of methods that read and write to {\tt f,g} are identified. The {\small \tt fieldReadWrites} procedure identifies these methods for a given field. The
 procedure receives a field and a class as input. It returns a list of methods in the class that read the field and a list of methods 
 that write to it. 
 
 For the {\tt ArrayList} example in Figure~\ref{fig:arraylist-repl}, invoking the procedure with the 
 field {\small \tt elementData} as an input will return the following lists: {\small \tt \{set,contains,indexOf\},
 \{set\}}. The {\small \tt indexOf,set} methods are added to the read list, as they both
 read {\small \tt elementData}.\footnote{{\tt indexOf} not shown for brevity}. 
 The {\small \tt contains} method is added to the list, because it 
 indirectly reads {\small \tt elementData} by invoking {\small \tt indexOf} method.
 A similar analogy is applied to identify the set of methods that
 write to {\tt elementData}.
 
 Once the read and write list for fields {\tt f,g}
 is identified, the similarity between the corresponding lists is computed using the {\tt similarity} function. The function creates vector embeddings for input lists and computes their cosine similarity score. {\tt FieldMapper} computes 
 the similarity score for both field 
 reads and writes (lines {\tt 7-8}). 
  The average of these scores is stored in {\tt escore}. The 
  weighted sum of {\tt escore} and
 the type similarity score for {\tt f,g} (stored in $TS$) is 
 added to matrix $FS$. The weight of the scores is determined by the  
 value of {\tt fw}, set by the user.
 
 After all the field pairs are processed, the matrix $FS$ contains the similarity scores for all field pairs. {\tt OPTIMIZE} uses $FS$ to compute the best match between the fields that maximizes 
 the overall score. We used the Hungarian Algorithm\cite{Hungarian} to 
 instantiate {\tt OPTIMIZE} function in our implementation, but any algorithm that finds an optimal
 match for a weighted bipartite graph can be used. The returned map is
 filtered by the {\tt FieldMapper} to eliminate pairs with low 
 scores. The $\mathtt{ft}$ value set by the user determines the
 minimum score for a field pair. The filtered map, $\sigma$, is returned to the main \tool{} algorithm, which is subsequently used by the {\tt MethodMapper}.

\subsection{Method Mapper}\label{sec:method-mapper}
The {\tt MethodMapper} receives classes \Q{}, {\tt R}, the field map
$\sigma$ and the type similarity matrix $TS$ as input. It creates 
a method map $\alpha$ based on the functional and type similarity between 
methods. Algorithm~\ref{alg:methodscore} presents the overall working of
{\tt MethodMapper}.

%

\begin{algorithm}[h]
	\footnotesize
	\caption{\label{alg:methodscore}$\mathtt{MethodMapper}$ algorithm}
	\begin{algorithmic}[1]
		\Require{\Statex {\tt Q} $\gets$ Query class; {\tt R} $\gets$ Replacement class
		\Statex $\sigma \gets$ $\mathtt{Field\ Map}$; ${TS} \gets$ Type similarity matrix
		\Statex $\mathtt{id} \gets$ Maximum in-line depths\Comment{Configuration variables}
        \Statex $\mathtt{mt} \gets$ Method similarity thresholds;  $\mathtt{mw} \gets$ Method embedding weight
		}
		\State ${MS} \gets \mathtt{INIT(Q, R)}$
		\State $\mathtt{R' \gets rewrite(R, \sigma)}$
		\For {every $\mathtt{(m_i, m_j)} \in \mathtt{methods(\texttt{Q})}\times \mathtt{methods(R')}$}
		\State $\mathtt{Tokens_i} \gets \mathtt{MethodTokenizer(m_i, id)}$;  
		\State $\mathtt{Tokens_j} \gets  \mathtt{MethodTokenizer(m_j, id)}$
		\State $\mathtt{emb\text{-}score} \gets \mathtt{similarity(Tokens_i, Tokens_j)}$
		\State ${PS} \gets \mathtt{INIT(m_i,m_j)}$
		\For {every $\mathtt{(p, q)} \in \mathtt{Parameter(m_i)}\times \mathtt{Parameter(m_j)}$}
		\State $PS[\mathtt{p,q}] \gets TS[\mathtt{type(p), type(q)}]$
		\EndFor
		\State $\mathtt{parameter\text{-}map} \gets \mathtt{OPTIMIZE}(PS)$;
		\State $\mathtt{par\text{-}score} \gets \mathtt{normalize}(\mathtt{parameter\text{-}map}, PS)$
		\State $MS[\mathtt{m_i,m_j}] \gets 
		\mathtt{mw}*\mathtt{emb\text{-}score}
		+ \left(1- \mathtt{mw}
		\right) * \mathtt{par\text{-}score}$
		\EndFor
		\State $\alpha \gets \mathtt{best\text{-}match}({MS}, \mathtt{mt})$
		\State \textbf{return} $\alpha$
	\end{algorithmic}
\end{algorithm}

The algorithm begins by creating a matrix $MS$, that will store the
similarity score for every method pair ($m_i,m_j$), where $m_i$ is defined by \Q{} and
$m_j$ is defined by {\tt R}. The 
matrix is initialized with {\tt -1} score for all pairs using the 
{\tt INIT} procedure. Next, the algorithm rewrites the class {\tt R}
by renaming fields based on the input field map, $\sigma$.
For the {\tt ArrayList} and {\tt Vector} classes given in Figure~\ref{fig:arraylist-repl}, $\sigma$ mapped {\tt size} field in {\tt ArrayList} to {\tt elementCount} field in {\tt Vector}. Therefore, 
{\tt MethodMapper} renamed the {\tt elementCount} field as {\tt size}.

After the rewrite, the algorithm iterates over all pairs of
{\em public} methods $({m_i,m_j})$ in \Q{}, {\tt R$'$}. 
Given a pair of methods $({m_i,m_j})$, the 
algorithm extracts tokens that capture the functionality of ${m_i,m_j}$, 
using {\tt MethodTokenizer} function. Before extracting the tokens, 
the procedure inlines
every method invocation within the input method. This is carried out upto a specific depth specified by the 
{\tt id} value provided by the user. 
We implemented {\tt MethodTokenizer} to extract the list of variable names and method names accessed by the method. This included the names of the invoked methods, the
parameter names, field names and the local variables defined by a method. However, {\tt MethodTokenizer} can be replaced with any function that 
can extract a list of tokens that represents the method functionality.
The tokens for the methods are embedded into vectors and similarity
between the vectors is computed by the {\tt similarity} 
function (line {\tt 6}). 

Next, the algorithm creates a map between the parameters received by ${m_i,m_j}$, based on their type similarity. 
{\tt MethodMapper} frames parameter mapping as an 
assignment problem\cite{assignmentproblem}. It creates
a matrix, $PS$, that stores type similarity
score for every pair ($\mathtt{p,q}$), where 
$\mathtt{p,q}$ are parameters to
${m_i,m_j}$ respectively (lines {\tt 8-10}).
This matrix is input to the {\tt OPTIMIZE} function, 
which returns an optimal map that maximizes the overall score. We instantiated {\tt OPTIMIZE} with
the Hungarian algorithm\cite{Hungarian}, but it can be replaced with any algorithm that solves the assignment problem.
The overall score of the match returned 
by the {\tt OPTIMIZE} function is normalized to fit within a range 
{\tt -1 to +1}. The normalized score is used as the parameter 
mapping score (line 12).
The final score assigned for the pair ${m_i,m_j}$ is the weighted
sum of the parameter mapping score and the method embedding score, which
will be stored in $MS$. The
weight for each score is determined by the {\tt mw} value, specified by the user. 

Once all pairs are processed, the algorithm scans the 
$MS$ matrix to identify the required $\alpha$. Each method ${m_i}$
in \Q{} is mapped to a method ${m_j}$ with the highest score. If the
highest ranking candidate is below a threshold {\tt mt} set by the user, the method will be
mapped to $\bot$. 

The newly constructed $\alpha$ is returned to the \tool{} top-level algorithm, which
ranks the classes based on the number of mappings in their $\alpha$. Ties 
are broken based on the aggregate method scores in $\alpha$. If ties persist, they are resolved based on the field mapping $\sigma$. The top-level algorithm returns the ranked classes and the mappings to the user.

\vspace{-0.1in}
\section{Experimental Results}
\label{sec:experiments}
To evaluate \tool{}, we formulated the following research questions:

\begin{itemize}[leftmargin=*]
  \item RQ1: Is \tool{} able to highly rank appropriate replacement classes?
  \item RQ2: Is \tool{} able to appropriately map methods in the original class
        to methods in the replacement class?
  \item RQ3: Does \tool{}'s hybrid methodology, which blends type analyses
        and embeddings-based search, improve on the class replacement task over just using
        embeddings?
\end{itemize}

To investigate these research questions, we carried out three different experiments. We now
outline the corresponding methodology.

\subsection{Methodology}
\begin{table}[htbp]
\footnotesize
  {\centering
  \caption{Benchmark Information. 
  |{\tt LOC}| is the lines of code. 
  |{\tt I}| is the number of classes explicitly imported by the class. |{\tt F}| and |{\tt M}| are field and method counts respectively. |{\tt P}| is the maximum parameters received by any method in the class. 
  }\label{tab:benchmarks}
 \begin{tabular}{|l|l|c|c|r|r|r|r|}
    \hline
    ID & Class name              & Package                                     & $|${\tt LOC}$|$ & $|${\tt I}$|$ & $|${\tt F}$|$ & $|${\tt M}$|$ & $|${\tt P}$|$   \\ \hline\hline
    1  & {\tt ArrayList}         & \multirow{2}{*}{java.util}                  & 1731            & 6             & 15            & 32            & 3             \\ \cline{1-2}\cline{4-8}
    2  & {\tt Vector}            &                                             & 1482            & 7             & 5             & 50            & 3            \\ \hline
    3  & {\tt FastArray}         & groovy.util                                 & 144             & 5             & 4             & 14            & 4              \\ \hline
    4  & {\tt FastVector}        & weka.core                                   & 216             & 3             & 4             & 47            & 3              \\ \hline
    5  & {\tt Box2}              & jmist.math                                  & 773             & 1             & 8             & 39            & 3             \\ \hline
    6  & {\tt Rectangle}         & eclipse.draw2d                              & 1414            & 1             & 6             & 88            & 5               \\ \hline
    7  & {\tt MutablePair}       & \multirow{3}{*}{apache.commons}             & 172             & 1             & 4             & 9             & 2                \\ \cline{1-2}\cline{4-8}
    8  & {\tt ImmutableTriple}   &                                             & 204             & 0             & 6             & 5             & 3                  \\ \cline{1-2}\cline{4-8}
    9  & {\tt MutableTriple}     &                                             & 172             & 0             & 5             & 9             & 3             \\ \hline
    10 & {\tt Point3DImpl}       & openimaj.math                               & 293             & 6             & 3             & 26            & 4             \\ \hline\hline
    11 & {\tt CircularFifoQueue} & \multirow{3}{*}{apache.commons} & 432             & 12            & 6             & 23            & 1               \\ \cline{1-2}\cline{4-8}
    12 & {\tt TreeList}          &                                             & 1132            & 11            & 3             & 19            & 3                \\ \cline{1-2}\cline{4-8}
    13 & {\tt FileEntry}         &                           & 276             & 5             & 10            & 17            & 2              \\ \hline
    14 & {\tt CacheStats}        & \multirow{2}{*}{google.common}              & 304             & 8             & 6             & 17            & 2             \\ \cline{1-2}\cline{4-8}
    15 & {\tt IntMath}           &                                             & 1350            & 18            & 0             & 24            & 3               \\\hline
    16 & {\tt XYDataItem}        & {jfree.data}                                & 240             & 3             & 3             & 11            & 2               \\\hline
    17 & {\tt BufferedReader}    & {java.io}                                   & 593             & 7             & 12            & 10            & 4               \\\hline
    18 & {\tt IndexedElement}    & dom4j.util                                  & 334             & 12            & 6             & 41            & 4               \\ \hline\hline
    19 & {\tt RequestUtil}       & \multirow{2}{*}{apache.sling}               & 104             & 3             & 0             & 4             & 4                \\  \cline{1-2}\cline{4-8}
    20 & {\tt ResponseUtil}      &                                             & 131             & 1             & 0             & 2             & 1               \\ \hline
    21 & {\tt TreeBagMultimap}   & eclipse.collections                         & 215             & 25            & 2             & 20            & 2                \\ \hline
    22 & {\tt SlowFuzzyQuery}    & apache.lucene                               & 201             & 10            & 7             & 6             & 2                 \\ \hline
    23 & {\tt Matrix}            & weka.core                                   & 558             & 3             & 2             & 27            & 4               \\ \hline
    24 & {\tt ManagedLinkedList} & groovy.util                                 & 147             & 3             & 7             & 4             & 2                \\ \hline
  \end{tabular}
  }
\vspace{-0.1in}
\end{table}

\subsubsection{Search Corpus and Query Classes}
To find replacement classes \tool{} relies on having access to a large
corpus of existing classes, from which it can draw candidate replacements. We constructed such a corpus for our
experiments. The corpus consists of the {\tt Java} classes previously used by Aroma~\cite{aroma},
{\tt Java} classes identified as 
popular~\cite{java-popular} and the classes used by prior work on synthesizing class replacements~\cite{mask}.
This corresponds to $\approx$600 thousand classes in the search database.

Given a search corpus, \tool{} finds a replacement class for the user's
original \emph{query} class. We collected a total of 24 classes for our queries.
Table~\ref{tab:benchmarks} provides an overview of these 24 query classes. The first three columns
correspond to a numeric identifier for the class (for brevity of discussion), the class name,
and the package from which it is sourced. The fourth column presents the number of lines of code
in the class. The fifth column presents the number of classes explicitly imported by the class. 
The sixth and seventh columns give the count of fields and public methods in the class respectively.
The maximum number of parameters received by any public method in the class is given in
the ninth column. 

Classes {\tt 1-10} are the set of classes used in prior work for synthesizing
replacement classes~\cite{mask}. This represent classes for which we know there exists potential replacement class  
in our search corpus. We refer to these classes as \emph{group 1}.
Classes {\tt 11-18} are additional classes that we found by inspecting libraries identified
as popular~\cite{java-popular}. We refined this to the set of classes that we could manually inspect
and determine whether replacement classes would be appropriate. This group of classes effectively represent
queries where we do not know if an appropriate replacement class exists in our search corpus.
We refer to these classes as \emph{group 2}.
Finally, classes {\tt 19-24} consists of classes that have been deprecated by software developers and
for which we have a ground truth on the class that was used to replace the original.
The packages containing the deprecated class and the replacement were part of the search corpus. 
We refer to these classes as \emph{group 3}.

\vspace{-0.1in}
\subsubsection{Configuration}\label{sec:config}
We set the type similarity threshold ($\mathtt{tt}$) to +0.8, 
field similarity threshold ($\mathtt{ft}$) to +0.5 and
method similarity threshold ($\mathtt{mt}$) to +0.5 in our experiments. 
These thresholds eliminate poor mappings and also speed up 
the optimization function: {\tt OPTIMIZE}. We set the field weight 
{\tt fw} and method weight to {\tt mw} to 
0.5. These values can range between 0-1.
Setting the weights to 0.5 informs \tool{} to give 
equal importance to type mapping and semantic equivalence. 
These can be modified based on the user's requirements.
Setting the value 0 will return classes that 
are type compatible but are semantically different. 
Setting it to 1 will yield results that are
semantically equivalent but are not type compatible.
The {\tt id} variable determines maximum method inlining depth and
it was set to 5. Increasing this
depth can improve the precision of method embeddings but will also 
incur a slowdown. We considered fields and methods in classes \Q{} and {\tt R} inherited 
from the parent classes, if any. This depth can also be defined by the user. 

\vspace{-0.1in}
\subsubsection{Searching and Analysis}
For each query class, we configured \tool{} to first retrieve
the top {\tt 1K} classes from the search database that are nearest to the query class, with the help of {\tt embedding} function. The distance between neighbors is defined to be the cosine similarity of their
corresponding class vectors (see section~\ref{sec:embedder} for details).
\tool{} analyzed all the {\tt 1k} classes to establish a mapping between the methods in the classes. 
Each class {\tt N} is assigned a rank based on the number of methods in {\tt Q} that are
mapped to a method in {\tt R} by {\tt ClassMapper}. The classes were sorted based on this rank and returned to the user. If two classes have the same rank, we use the number of mapped fields and the average method scores as a tie-breakers.

\vspace{-0.1in}
\subsubsection{Evaluation}
\label{subsubsec:evaluation}
To determine the performance of \tool{} we consider two approaches. In settings where we
have the ground truth replacement class (as determined by human developers), such as is
the case for the deprecated classes (19 -- 24), we use the rank assigned by \tool{} as
a metric of performance. A higher rank for the known ground truth corresponds to a better performance.
For cases where we do not have a ground truth replacement class and furthermore it is not necessarily
the case that our corpus contains an appropriate replacement class, we defer to manual inspection.
In particular, we retrieve the top 5 classes suggested by \tool{} and then manually inspect their source
code and compare this to the original query class to determine if they constitute valid replacements.
We perform a similar analysis for classes retrieved using only an embeddings-based search.
We compute the fraction of methods in the original class that can be replaced by
methods in the replacement class candidate in a 1-to-1 mapping. We also consider a nuanced breakdown
of the correctness criteria for a mapping.

To determine the quality of the method mapping ($\alpha$) 
between the original and replacement classes
produced by \tool{}, we assign each individual mapping to one of five possible quality categories: \{{\tt C1, C2, E1, E2, E3}\}. 
We now define each of these categories.

Given a manually determined ideal field mapping ($\sigma$) between class {\tt R} and {\tt Q}, the automatically generated method
mapping produced by \tool{} ($\alpha$), and a particular method ($m$) in the query class, we define the following categories:
\begin{itemize}[leftmargin=*]
      \item C1: a mapping for $m$ exists in the ideal mapping, and \tool{} maps it to the same method as the ideal.
      \item C2: no mapping exists for $m$ in the ideal mapping, and \tool{} maps it to empty.
      \item E1: a mapping for $m$ exists in the ideal mapping, but \tool{} maps it to the wrong method.
      \item E2: no mapping exists for the method in the ideal mapping, but \tool{} maps it to a method.
      \item E3: a mapping for $m$ exists in the ideal mapping, but \tool{} maps it to empty.
\end{itemize}


For completeness, we often refer to the fraction of public methods in the query class that can be correctly replaced by the replacement class. This corresponds
to the fraction of methods in $\alpha$ labeled as C1, E1, or E3.
We denote this fraction of methods as $\mathcal{P}$. The motivation behind this metric is to
draw a distinction between the quality of the mapping produced by \tool{} and the ranking
of replacement classes produced by \tool{}. We measure the quality
of method mapping with $\mathcal{C}$, this corresponds to the 
fraction of methods labeled C1 or C2.

To determine the quality of replacement classes' ranking, we compare the rank assigned
by \tool{} and the initial rank assigned by 
the {\em embedding} function in {\tt N}. We denote the ranking from the embedding-only approach as {\tt ER} (short for embedding rank).
This method constitutes the first part of the \tool{} pipeline, before method and field maps are created for the candidate replacement classes.

\noindent{\textbf{Evaluation Hardware}}
All the experiments are run on a Ubuntu 16.04 machine, running on a 3GHz Intel(R) Xeon(R) processor with 528 GB RAM and 40 cores.

\subsection{Results}
\begin{table*}
  \footnotesize
  \caption{Ranking results for query classes in \emph{group 1} and \emph{group 2}.
  For each query class (Class name), we present statistics for the top 5 \tool{} results.
  The $\mathcal{P}_e$, $\mathcal{P}$ corresponds 
  to the percentage of methods that can be ideally replaced by the top $n^{th}$ replacement class returned by just
  the  embedding-based search and the \tool{} respectively. The |{\tt ER}| is the initial rank assigned to the class by the
  embeddings-based search, before being re-ranked to the $n^{th}$ position by \tool{}'s complete pipeline. $\mathcal{C}$ is the precision of 
    the method map $\alpha$. 
    ${\tt Time}$ is the 
  time taken by \tool{} in minutes.
  }\label{tab:search-results}
  
  \begin{tabular}{|l
  |r|r|r|r|
  |r|r|r|r|
  |r|r|r|r|
  |r|r|r|r|
  |r|r|r|r|
  r|}
    \hline
    \multirow{3}{*}{Class name} & \multicolumn{4}{c||}{Rank 1} & \multicolumn{4}{c||}{Rank 2} & \multicolumn{4}{c||}{Rank 3} & \multicolumn{4}{c||}{Rank 4} & \multicolumn{4}{c|}{Rank 5} &             \multirow{3}{*}{Time} \\ \cline{2-21}
                                & {\tt N}   
                                & \multicolumn{3}{c||}{
                                \tool{}}                                         
                                & {\tt N}
                                & \multicolumn{3}{c||}{\tool{}}                                                
                                & {\tt N}
                                & \multicolumn{3}{c||}{\tool{}}                                     
                                & {\tt N}   
                                & \multicolumn{3}{c||}{\tool{}}                                     
                                & {\tt N}   
                                & \multicolumn{3}{c|}{\tool{}}
                                &  \\\cline{2-21}

    &$\mathcal{P}_e$            & {\tt ER}               & $\mathcal{P}$            & $\mathcal{C}$                                  
    &$\mathcal{P}_e$            & {\tt ER}               & $\mathcal{P}$            &$\mathcal{C}$                                     
    &$\mathcal{P}_e$            & {\tt ER}               & $\mathcal{P}$            &$\mathcal{C}$                                    
    &$\mathcal{P}_e$            & {\tt ER}               & $\mathcal{P}$            &$\mathcal{C}$                                     
    &$\mathcal{P}_e$            & {\tt ER}               & $\mathcal{P}$            &$\mathcal{C}$        &  \\\hline

{\tt ArrayList}     &100    & 366   & \textbf{100}   & 100     
                    &100    & 462   & \textbf{100}   &100
                    & 87    & 959   & \textbf{100}   & 100  
                    &0      & 675   & \textbf{100}   & 100     
                    &0      & 124   & \textbf{100}   & 84 & 90     \\ \hline
{\tt Vector}        &79     & 78    & \textbf{91}    & 83  
                    &79     &4      & \textbf{83}    & 83 
                    &0      & 2     & \textbf{87}    & 95    
                    &69     & 606   & \textbf{73}    &70
                    &0      & 1     & \textbf{79}    &81   & 100      \\\hline
{\tt FastArray}     & 50    & 979   & \textbf{57}    &57                       
                    &50     & 64    & \textbf{64}    &64                      
                    &0      & 69    & \textbf{57}    &57          
                    &7      & 253   & \textbf{50}    & 64          
                    &50     & 96    & \textbf{42}    &71 &42     \\\hline
{\tt FastVector}    &68     & 104   & \textbf{85}    & 91                      
                    &17     & 194   & \textbf{85}    & 91                      
                    &40     & 70    & \textbf{85}    & 91         
                    &21     & 43    & \textbf{72}    & 79         
                    &91     & 6     & \textbf{91}    & 85   &95       \\ \hline
{\tt Box2}          & 76    & 1     & \textbf{76}    & 84
                    & 0     & 156   & \textbf{29}    & 84                      
                    & 0     & 284   & \textbf{30}    & 71         
                    &33     & 15    & \textbf{30}    & 74         
                    & 33    & 12    & \textbf{33}    & 94  &91     \\ \hline
    
{\tt Rectangle}     & 100   & 1     & \textbf{100}   & 100                     
                    & 42    & 80    & \textbf{46}    & 72                      
                    & 34    & 2     & \textbf{42}    & 70         
                    & 0     & 27    & \textbf{37}    & 67         
                    & 34    & 6     & \textbf{37}    & 70  &61     \\ \hline
{\tt MutablePair}   & 62    & 1     & \textbf{62}    & 75                      
                    & 75    & 2     & \textbf{75}    & 87                  
                    & 0     & 11    & \textbf{75}    & 87         
                    & 50    & 7     & \textbf{75}    & 87         
                    & 50    & 10    & \textbf{62}    & 75   &13      \\ \hline
{\tt ImmutableTriple}& 50   & 2     & \textbf{66}    & 100                     
                    & 66    & 3     & \textbf{83}    & 83                      
                    & 83    & 1     & \textbf{50}    & 83         
                    & 0     & 5     & \textbf{33}    & 83         
                    & 33    & 20    & \textbf{50}    & 100  &7      \\ \hline
{\tt MutableTriple} & 0     & 2     & \textbf{50}    & 75                      
                    & 50    & 8     & \textbf{87}    & 87                      
                    & 62    & 7     & \textbf{75}    & 75         
                    & 50    & 6     & \textbf{75}    & 75         
                    & 25    & 19    & \textbf{50}    & 62  &14     \\ \hline
{\tt Point3DImpl}   & 19    & 5     & \textbf{30}    & 69                  
                    & 11    & 3     & \textbf{30}    & 69                  
                    & 30    & 1     & \textbf{19}    & 57    
                    & 11    & 2     & \textbf{11}    & 57    
                    & 30    & 4     & \textbf{11}    & 53  &66     \\ \hline
{\tt CircularFifoQueue}&100 & 1     & \textbf{100}   & 100                  
                    & 42    & 51    & \textbf{86}    & 65                 
                    & 42    & 119   & \textbf{86}    & 73    
                    & 42    & 106   & \textbf{86}    & 86    
                    & 17    & 133   & \textbf{86}    & 73  & 33    \\ \hline
{\tt TreeList}      & 57    & 8     & \textbf{100}   & 100                  
                    & 57    & 5     & \textbf{100}   & 100                
                    & 57    & 81    & \textbf{100}   & 94    
                    & 100   & 309   & \textbf{100}   & 100    
                    & 84    & 423   & \textbf{100}   & 100  &30     \\ \hline
{\tt FileEntry}     & 11    & 8     & \textbf{58}    & 70                  
                    & 29    & 18    & \textbf{64}    & 94                 
                    & 23    & 286   & \textbf{58}    & 82    
                    & 47    & 7     & \textbf{41}    & 70    
                    & 41    & 29    & \textbf{52}    & 76  &23     \\ \hline
{\tt CacheStats}    & 100   & 1     & \textbf{100}   & 100                  
                    &23     & 18    & \textbf{58}    & 47                 
                    &29     & 6     & \textbf{11}    & 47   
                    &29     & 16    & \textbf{23}    & 70    
                    &0      & 577   & \textbf{5}     & 52  &11    \\ \hline
{\tt IntMath}       &100    & 3     & \textbf{100}   & 100              
                    &100    & 2     & \textbf{100}   & 81                 
                    &100    & 1     & \textbf{100}   & 95   
                    &45     & 19    & \textbf{36}    & 59    
                    &18     & 69    & \textbf{59}    & 31  &23     \\ \hline
{\tt XYDataItem}    &0      & 261   & \textbf{72}    & 90                      
                    &0      & 34    & \textbf{54}    & 72                 
                    &0      & 192   & \textbf{63}    & 72    
                    &72     & 160   & \textbf{54}    & 72    
                    &0      & 27    & \textbf{45}    & 63  &17    \\ \hline
{\tt BufferedReader}&0      & 7     & \textbf{100}   & 100                  
                    &0      & 65    & \textbf{60}    & 60                 
                    &100    & 239   & \textbf{70}    & 70    
                    &0      & 33    & \textbf{90}    & 90   
                    &0      & 6     & \textbf{100}   & 90  &46    \\ \hline
    
{\tt IndexedElement}&0      & 283   & \textbf{95}    & 97                      
                    &0      & 38    & \textbf{95}    & 100                     
                    &0      & 24    & \textbf{47}    & 82         
                    &0      & 340   & \textbf{17}    & 87         
                    &0      & 95    & \textbf{17}    & 87  & 48     \\ \hline \hline
    
Average             &54     & 117   & \textbf{80}    &  88                    
                    &41     & 67    & \textbf{72}    &  79                     
                    &38     & 130   & \textbf{64}    &  77        
                    &32     & 146   & \textbf{55}    &  77        
                    &28     & 94    & \textbf{56}    &  74 &45m   \\ \hline
  \end{tabular}
  \hspace{0.3cm}
\end{table*}

Table~\ref{tab:search-results} presents the results of our search experiments for classes
in group 1 and group 2.
The first column presents the name of the query class. The columns {\tt 2-15} present statistics
on the top 5 classes
returned by the intermediate embedding-based search and those returned by \tool{}. The columns titled $\mathcal{P}_e$ presents details 
about the top ranked class in the intermediate list {\tt N} ({\tt ER rank 1-5}). The 
column presents the percentage of methods that
can be correctly replaced by this class and is used to measure the quality of classes
if method and field mapping were skipped.
The columns titled $\mathtt{ER}, \mathcal{P}$ and $\mathcal{C}$ provide
details about the final top ranked replacement class returned
by \tool{}. This is the class field and method mappings placed at the top. The $\mathtt{ER}$ presents the
initial rank of this class in {\tt N} and $\mathcal{P}$
provides the percentage of query class methods it can 
replace. For example, for {\tt Vector} class, $\mathcal{P}_e$ notes that the
top replacement class produced by the embeddings-based search (i.e. the first
class in {\tt N}) can correctly replace
  {\tt 79\%} of {\tt Vector} class's methods. In contrast, the top result for
\tool{} can correctly replace {\tt 91\%} ($\mathcal{P}$).
$\mathcal{C}$ presents the accuracy of method mapping generated
by \tool{}. This includes methods it identified correct replacement
or correctly identified the absence of one.

On average, the top ranked classes identified by \tool{} have a corresponding method replacement
for 80\% of the methods in the original class (with multiple cases of 100\% coverage). In contrast,
the top classes retrieved solely based on an embeddings search can replace an average 54\% methods ($\mathcal{P}_e$). 
This trend continues to manifest across all top 5 rankings. 
Furthermore, \tool{}'s ranking provides substantial refinement over the original
ranking produced by the embedding-only approach. In particular, for 6 of the query classes
the top ranked class returned by \tool{} had a rank between 78 and 979 in the
embeddings-only approach.



\begin{figure*}[h]
  \begin{subfigure}{0.24\textwidth}
    {\includegraphics[width = 1.8in]{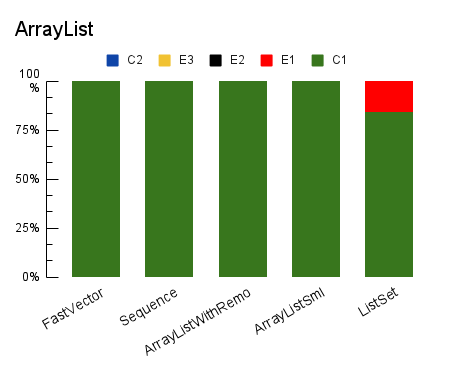}}
  \end{subfigure}%
  ~
  \begin{subfigure}{0.24\textwidth}
    {\includegraphics[width = 1.8in]{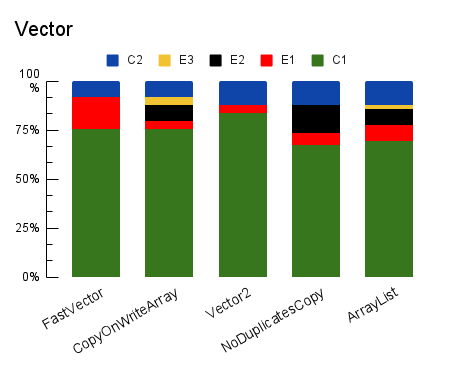}}
  \end{subfigure}
  \begin{subfigure}{0.24\textwidth}
    {\includegraphics[width = 1.8in]{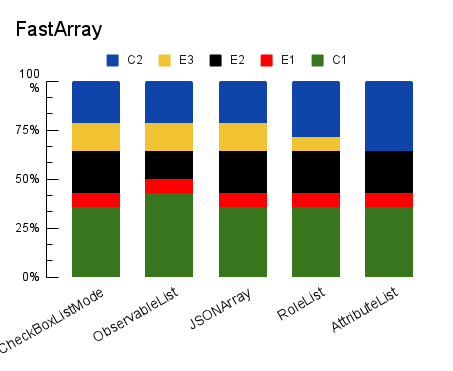}}
  \end{subfigure}%
  ~
  \begin{subfigure}{0.24\textwidth}
    {\includegraphics[width = 1.8in]{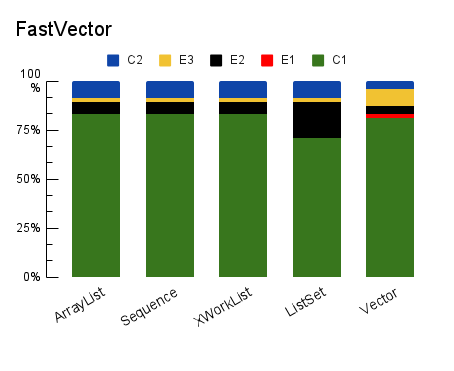}}
  \end{subfigure}
  \\
  \begin{subfigure}{0.24\textwidth}
    {\includegraphics[width = 1.8in]{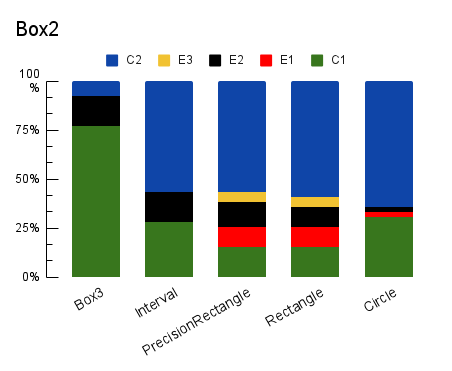}}
  \end{subfigure}%
  ~
  \begin{subfigure}{0.24\textwidth}

    {\includegraphics[width = 1.8in]{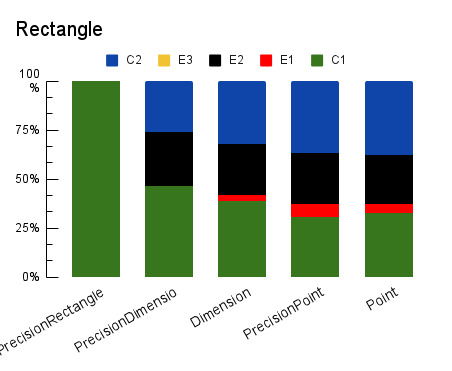}}
  \end{subfigure}
  ~
  \begin{subfigure}{0.24\textwidth}
    {\includegraphics[width = 1.8in]{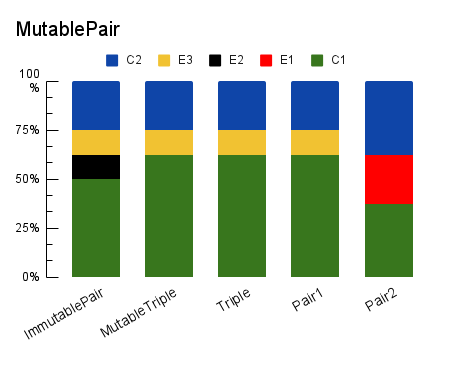}}
  \end{subfigure} %
  ~
  \begin{subfigure}{0.24\textwidth}
    {\includegraphics[width = 1.8in]{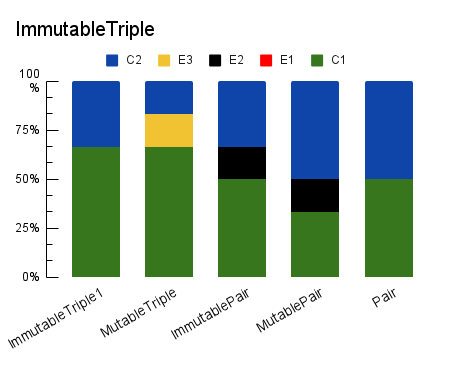}}
  \end{subfigure}
  \\
  \begin{subfigure}{0.24\textwidth}
    {\includegraphics[width = 1.8in]{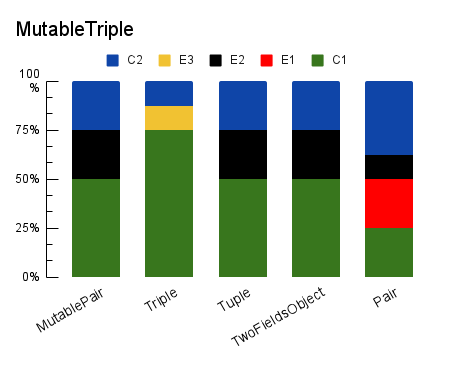}}
  \end{subfigure} 
  ~
  \begin{subfigure}{0.24\textwidth}
    {\includegraphics[width = 1.8in]{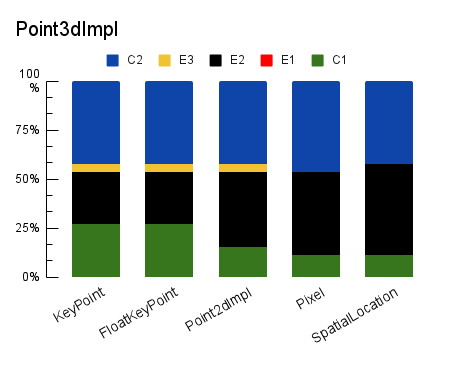}}
  \end{subfigure}
  ~
  \begin{subfigure}{0.24\textwidth}
    {\includegraphics[width = 1.8in]{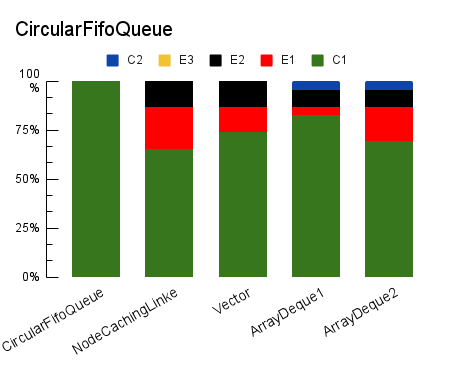}}
  \end{subfigure} %
  ~
  \begin{subfigure}{0.24\textwidth}
    {\includegraphics[width = 1.8in]{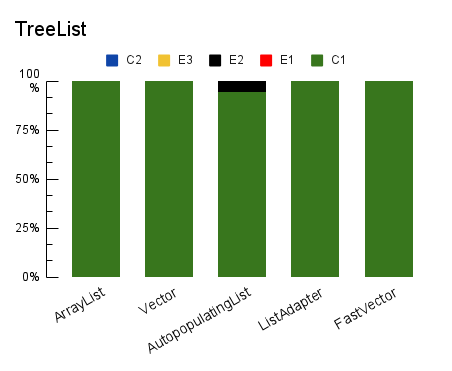}}
  \end{subfigure}\\
  \begin{subfigure}{0.24\textwidth}
    {\includegraphics[width = 1.8in]{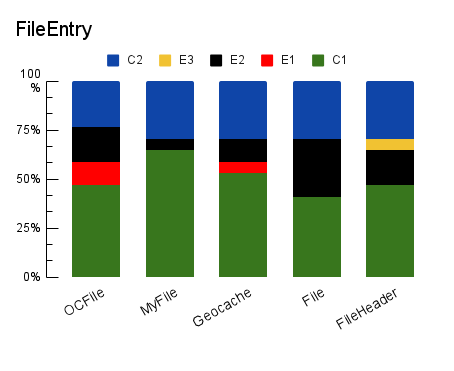}} \end{subfigure} %
  ~
  \begin{subfigure}{0.24\textwidth}
    {\includegraphics[width = 1.8in]{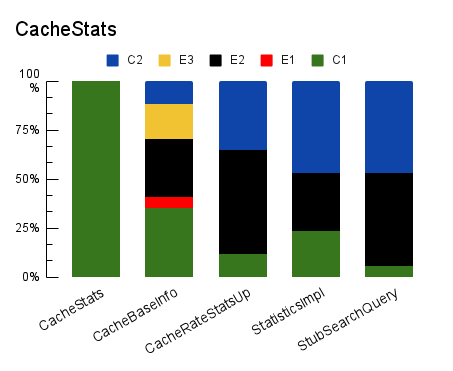}}
  \end{subfigure}
  ~
  \begin{subfigure}{0.24\textwidth}
    {\includegraphics[width = 1.8in]{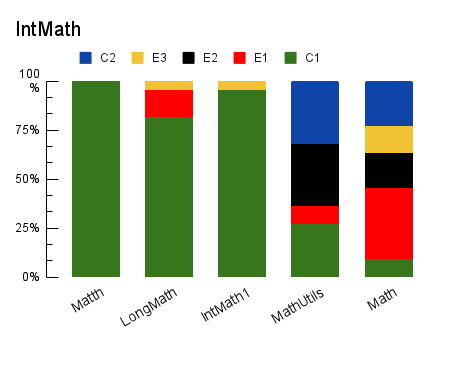}}
  \end{subfigure}
  ~
  \begin{subfigure}{0.24\textwidth}
    {\includegraphics[width = 1.8in]{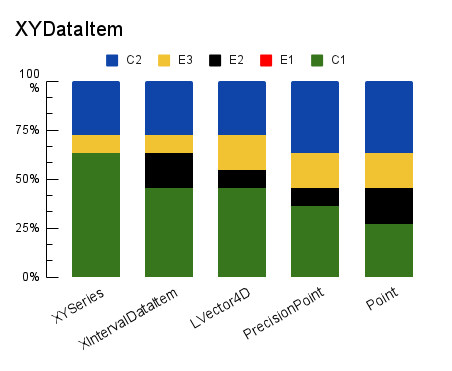}}
  \end{subfigure}\\
  \begin{subfigure}{0.24\textwidth}
    {\includegraphics[width = 1.8in]{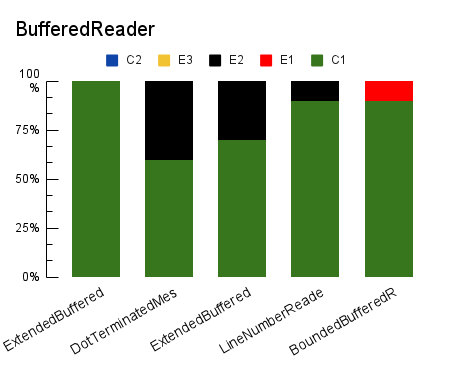}}
  \end{subfigure}%
  ~
  \begin{subfigure}{0.24\textwidth}
    {\includegraphics[width = 1.8in]{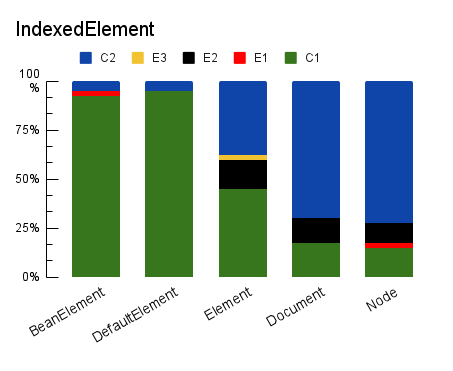}}
  \end{subfigure}
  \vspace{-0.1in}
  \caption{
    The quality classification of methods for the $\alpha$ constructed by the \tool{}.
    Subplot titles correspond to query classes. X-axis labels correspond to the top 5
    \tool{} results. The y-axis corresponds to the fraction of methods in the original
    class, and the colors denote the fraction of such methods that are associated with
    a given method mapping quality classification. C1 + C2 constitute correct mappings.
  }
  \label{fig:classification}
\end{figure*}

Next we present our results on the correctness of the method mappings produced by
\tool{}. The $\mathcal{C}$ values in Table~\ref{tab:search-results} demonstrate the average accuracy of method mappings is above 70\% for top 5 ranking classes. In some cases, the accuracy is 100\%. A more
detailed analysis of the precision is presented in Figure~\ref{fig:classification}.

Figure~\ref{fig:classification} presents a breakdown of the classification
of mappings for top ranked replacement candidates. Specifically, the title of each subplot
corresponds to the query class we want to replace. The x axis indicates the name of the top 5 ranked
replacement classes produced by \tool{}. The y axis denotes the fraction of methods in the query
class that belong to a method mapping quality category (see Subsection~\ref{subsubsec:evaluation})
and the colors that constitute each bar correspond to the quality categories.
Our results show that a significant fraction of method mappings produced by \tool{}
are correct (C1 + C2). As expected, this fraction is higher for classes that are more highly
ranked, though we observe some variation depending on the query class. And this variation can
represent a complex relationship between query class, replacement availability, and quality
of the mappings produced.

Finally, we present \tool{}'s results when queried with classes in \emph{group 3}, which
correspond to deprecated classes where we know exactly what class the developers used as a
replacement in their codebase. Given that we have the ground truth, and we know these
replacement classes exist in our class corpus, we evaluate the rank assigned to that
ground truth class as well as the method mapping produced.
Table \ref{tab:dep-results} presents the deprecated class name, information on
the package and naming of the replacement class, the rank produced by \tool{} for the ground truth
replacement,
the rank produced by the embeddings-only approach for the ground truth replacement, the
fraction of methods in the query class that can be correctly replaced ($\mathcal{P}$)
and a summary of the correct mappings produced by \tool{} (C1 + C2).
In all cases, \tool{} correctly retrieves the replacement class as the
top result. In contrast, the embeddings only approach fails for the
  {\tt Matrix} class, where it ranks the ground truth at 241.
Additionally, \tool{} does not just produce a ranking of replacement class, but also
a mapping between methods. We find that a large fraction (between 74\% and 100\%)
of \tool{}'s mappings for these classes' methods are correct.
It is worth noting that the ground truth replacement classes for two of the deprecated
classes ({\tt SlowFuzzyQuery} and {\tt Matrix}) are in fact not perfect replacements
(as demonstrated by the low $\mathcal{P}$). However, in such cases \tool{}'s mapping
still achieves a high fraction of correct mappings (C1 + C2) as it can appropriately
identify that some methods are not replaceable (C2) with the given replacement class.


\begin{table}[htbp]
\footnotesize
  {\centering
    \caption{Deprecated class results. PN and CN indicate if both the original and ground-truth replacement classes
      are in the same package and have the same name respectively. {\tt ER} indicates the initial rank
      assigned by the embedding based search. {\tt Final} is the final rank assigned by \tool{}. $\mathcal{P}$ provides the percentage of methods in the deprecated class that can be replaced by the top \tool{} result. $\mathcal{C}$ is the precision of 
    the method map $\alpha$.}\label{tab:dep-results}
    \begin{tabular}{|l|c|c|c|c|c|c|}
      \hline
      \multirow{2}{*}{Deprecated class}
      & \multirow{2}{*}{PN} & \multirow{2}{*}{CN} & \multicolumn{2}{c|}{Rank}& \multirow{2}{*}{$\mathcal{P}$} & \multirow{2}{*}{$\mathcal{C}$}  \\\cline{4-5}
                           & &  & {\tt ER} & {\tt Final} 
                           &  & \\\hline
      RequestUtil       & No           & Yes       & 1          & 1           & 100\%         & 100\%                                    \\ \hline
      ResponseUtil      & No           & Yes       & 1          & 1           & 100\%         & 100\%                                   \\ \hline
      TreeBagMultimap   & No           & Yes       & 1          & 1           & 100 \%        & 100\%                                     \\\hline
      SlowFuzzyQuery    & No           & No        & 1          & 1           & 83\%          & 100\%                                       \\\hline
      Matrix            & No           & Yes       & 241          & 1         & 37 \%         & 74\%                                     \\\hline
      ManagedLinkedList & No           & No        & 1          & 1           & 100\%         & 100\%                                     \\\hline
    \end{tabular}
  }
\end{table}

\subsection{Research Question Discussion}

We introduced this evaluation\footnote{
The data used in our experiments and the search results are available for download. The data will be publicly available upon acceptance.} with three key research questions, we now provide discussion of these based on our results.

\subsubsection{Research Question 1}
Our results show that \tool{} is capable of ranking appropriate replacement classes highly.
In particular, we found that for query classes in \emph{group 1} and \emph{group 2}
the top ranked \tool{} replacement class could be used to replace on average 80\% of the
methods in the original class (see Table \ref{tab:search-results}). The second and third ranked results could
still on average cover 72\% and 64\% respectively.  When we considered the case of query classes
from our \emph{group 3}, where we know the developer's ground truth replacement as the original
class was deprecated, we found that \tool{} produced the ground truth replacement as its
top result in all 6 cases.

\subsubsection{Research Question 2}
A key contribution of \tool{} is to produce not just a ranking of replacement classes
but also a detailed mapping between methods and fields in the original and replacement classes.
For our evaluation, we focused on the method mapping.
For classes in \emph{group 1} and \emph{group 2} we found that \tool{}'s mapping for the top ranked result
had a high fraction of correct mappings (C1 + C2) for most query classes in our evaluation.
For 11 of theses classes over 75\% of methods were correctly mapped.
Similarly, we found that for the deprecated classes (\emph{group 3}) \tool{}'s mappings
are correct in almost all cases. The one exception was {\tt Matrix}, where the replacement
class does not have a lot of the necessary functionality, and \tool{} incorrectly maps
26\% of methods (E1 + E2 + E3).

\subsubsection{Research Question 3}
Our results show that the additional analyses result in better replacement
rankings. In particular, for deprecated classes, we found that an embeddings-only approach
failed to retrieve the ground truth for the {\tt Matrix} class. Similarly, we showed that
the top ranked \tool{} results for group 1 and group 2 classes could have substantially different
rank in the embeddings-only approach. On average, the top ranked result produced by \tool{} can correctly
replace {\tt 80\%} of public methods, while the top ranked result produced by an embeddings-only
approach can correctly replace {\tt 54\%} of public methods. This relationship holds true
for the top 5 results: \tool{}'s re-ranking yields classes that can correctly replace more public
methods than the result in the same position produced by an embeddings-only approach.
Additionally, in contrast to an embeddings-only approach \tool{} can produce
not just a ranking but also a field and method mapping between the original and replacement class.

\section{Limitations and Threats to Validity}
\label{sec:threats}
The quality and the number of replacement classes identified by \tool{} depends
on the query class and the search database. For example, in
our evaluation, \tool{} did not identify a complete replacement for
{\tt FastArray}. This is a result of the unique implementation of {\tt FastArray}, as
it assumes the user will perform safety checks (e.g. bound checking arrays). In
contrast, most classes in the search database include classes that already
incorporate such checks. This mismatch in functionality, results in few classes
that can correctly (and directly) replace {\tt FastArray}.
To mitigate this limitation, a user should aim to collect a large enough
corpus of classes to populate \tool{}'s search database.

A key contribution in \tool{} is the combination of an embeddings-based search
and a type-aware mapping of candidate replacement classes. While type information
can improve \tool{}'s ability to retrieve appropriate candidates, it may also
lead to over constraining results. For example, in our evaluation the
    {\tt Box2} class uses a {\tt double} type to store rectangle coordinates.
In contrast, most rectangle class implementations in our search space
use less precise datatypes such as {\tt float} to represent these coordinates.
Since a {\tt double} cannot be represented safely as {\tt float} in all cases
(without additional, and potentially incomplete, analysis), \tool{} found few candidate
replacements for {\tt Box2}. To mitigate this limitation, a user may consider
adjusting the types in their query class, based on their knowledge of their application,
to increase the possible types that can safely represent their values.

\tool{}, and our evaluation, are focused on identifying drop-in replacement
classes. This can be a restrictive condition, and there may exist other classes
that can replace the original class' functionality, but require the user to
implement adapter code. Indeed, prior work~\cite{mask} showed that given an
original and a replacement class, automatically synthesizing adapter code is
feasible. To cover such cases, a user could consider lower ranked \tool{}
results as possible adapter-based replacement classes, where possible.

Different query classes may be more amenable to replacement and may
produce different candidate replacements when queried through \tool{}.
To mitigate this risk in our evaluation, we considered a total of 24 classes,
drawn from various sources. 

\tool{} uses embeddings to perform an initial search for candidate replacement
classes and to compute scores for possible field/method mappings during later
phases. Different ways of computing embeddings may produce different results. To
mitigate this risk, we chose to implement \tool{} in a modular fashion, where
different embedding functions can be used. We evaluated a version of \tool{}
that uses FastText~\cite{bojanowski2016enriching} to compute embeddings for
replacement classes, and their methods/fields. FastText is a popular system and
has previously been used within the context of code
search~\cite{cambronero2019deep}. Users could consider trying alternative versions
of \tool{} where the embedding function is instantiated differently, as a way of
diversifying the resulting candidate replacement classes.

\vspace{-0.15in}
\section{Related Work}\label{sec:related}

\noindent \textbf{API migration and class replacement} The task of automatically translating or replacing portions of an existing
program to use a new API has been explored in prior research. Zhong et
al~\cite{zhong2010mining} use a parallel corpus of projects (e.g. in two
languages) to mine API mappings by analyzing class usage in the paired examples.
Samak et al~\cite{mask} developed Mask, a system that combines symbolic
execution and synthesis to replace a given existing class with a given target
class. Nguyen et al~\cite{nguyen2014statistical}'s StaMiner mines mappings
between two APIs by aligning API call sequences in parallel client code. Given
paired client code, StaMiner aligns sequences of API calls, and identifies often
recurring alignments as likely mappings. Ni et al~\cite{ni2021soar} developed
SOAR, which automatically rewrites a data science program from one API to
another by combining information from the APIs' documentation, error traces from
partial executions, and enumeration-based synthesis. In contrast, \tool{} is not designed for synthesizing replacements based on two known libraries, it does not require a
parallel corpus of code which can be expensive to collect, and it produces
class-level mappings that associated methods and fields in the original and
replacement class.

\noindent \textbf{Neural code representations} Learning vector representations for code has become an active area of research.
Different approaches vary in the way they compute embeddings and the tasks that
their method is focused on solving. Alon et al~\cite{alon2019code2vec} extracts
vector representations for methods by combining embeddings produced from paths
through the method's AST. Henkel et al show that these embeddings can be used
for method naming, API mapping, and bug finding. Allamanis et
al~\cite{allamanis2018learning} introduced the use of Graph Neural Networks
(GNN) for learning code embeddings. Wang et al~\cite{wang2020learning}
introduced the Graph Interval Neural Network, which learns over an abstracted
representation of the program. Recently, large pre-trained transformer-based
language models have also been applied to code and related tasks such as program
synthesis, prominent examples include CodeBERT~\cite{feng2020codebert} and
Codex~\cite{chen2021evaluating}. \tool{} does not introduce a new way of
computing vector representations for code. In particular, while \tool{} relies
on FastText~\cite{bojanowski2016enriching} to compute vector representations,
the system is designed such that we could replace this with the techniques
presented previously. A key contribution is to show that we can
combine vector representations for code with combinatorial optimization in a
design guided by considering key program properties such as types.

\noindent{\textbf{Semantic code search and code clone detection}}
Searching for code, given a large corpus of potential matches, has been actively
explored and continues to be a source of novel techniques and systems in the
software engineering community. \emph{Semantic code search} in particular is
meant to enable search over semantic properties of code, such as the expected
behavior, in contrast to techniques that focus exclusively on the use of
syntactic properties (e.g. term overlap). For example, Gu et
al~\cite{gu2018deep} and Cambronero et al~\cite{cambronero2019deep} use neural
networks to  produce code and natural language embeddings that can be used to
search a large corpus of methods given a natural language description of the
desired functionality. Premtoon et al's Yogo~\cite{premtoon2020semantic} system
represents a code fragment using its corresponding dataflow graph, as well as
rewrites of the graph based on a set of rules, and perform queries as searches
over the graphs. David and Yahav~\cite{david2014tracelet} also use a
rewriting-based strategy to enable code search, but their system is focused on
searching for function usage in a corpus of executables. Closely related to
semantic code search is research focused on code clone detection. There is a
large body of work on code clone detection using varied methods, including
neural networks~\cite{white2016deep, wang2020detecting}, code rewrite
rules~\cite{kamiya2002ccfinder}, and efficient detection of term and structural
overlap~\cite{jiang2007deckard, yuan2012boreas, sajnani2016sourcerercc}, among
others. In contrast, \tool{} is designed to retrieve classes that can
be used to \emph{replace} any functionality exposed in the original.

\section{Conclusion}\label{sec:conclusion}
We proposed a technique that receives
a query class \Q{} and a search corpus S, and a searches for classes in S that can replace \Q{}. \tool{} integrates embedding
based search with type awareness to refine the final search results. We demonstrate the effectiveness of \tool{} by searching in a corpus of $\approx$ 600K classes. 
\bibliographystyle{IEEEtran}
\bibliography{main}
\end{document}